\documentstyle[epsfig,twocolumn,aps,amssymb]{revtex}
\begin{document}
\draft

\title{Dissipation and Vortex Creation in Bose-Einstein Condensed Gases}
\author{B. Jackson,$^1$ J. F. McCann,$^2$ and C. S. Adams$^1$}
\address{$^1$Department of Physics, Rochester Building, University of Durham, 
South Road, Durham, DH1 3LE, UK.}
\address{$^2$Department of Applied Mathematics and Theoretical Physics,
Queen's University, Belfast, BT7 1NN, UK.}
\date{\today}
\maketitle
\begin{abstract}
We solve the Gross-Pitaevskii equation to study energy transfer
from an oscillating `object' to a trapped Bose-Einstein condensate.
Two regimes are found: for object velocities below a critical value, energy is
transferred by excitation of phonons at the motion
extrema; while above the critical velocity, energy transfer is {\it via}\ 
vortex 
formation. The second regime corresponds to significantly 
enhanced heating, in agreement with a recent experiment.    
\end{abstract}

The existence of a critical velocity for dissipation is central to the issue
of superfluidity in quantum fluids. The concept was first 
introduced by
Landau in his famous criterion \cite{khalatnikov}, where elementary 
excitations are produced 
above a velocity $v_L$. In liquid $^4{\rm He}$ this process refers to
the excitation of rotons, with $v_L \simeq 58\, {\rm ms}^{-1}$.
However, much smaller critical values are observed experimentally, which
prompted Feynman to propose that quantized vortices may be responsible 
\cite{feynman}.

Vortex nucleation in superfluid $^4 {\rm He}$ is difficult to explain
quantatively. Strong interactions within the liquid, plus thermal and quantum 
fluctuations, impede formulation of a satisfactory microscopic theory.
In contrast, Bose-Einstein condensation (BEC)
in trapped alkali gases \cite{ketterle99,dalfovo99} provides 
a relatively simple system for exploring superfluidity. 
Weakly-interacting condensates can be produced with a negligibly small 
non-condensed component. This allows an accurate description by a nonlinear 
Schr\"{o}dinger equation, often known as the Gross-Pitaevskii (GP) equation. 
The system also offers excellent control over
the temperature, number of atoms and interaction strength, as well as allowing
manipulation of the condensate using magnetic and optical forces 
\cite{ketterle99}.

Recent experiments have produced vortices by coherent excitation 
\cite{matthews99} and cooling of a rotating cloud \cite{madison99}. The 
existence of
vortices was also inferred by Raman {\it et al.} \cite{raman99}, where a 
condensate was probed
by an oscillating laser beam blue-detuned far from atomic resonance. The 
optical dipole force expels atoms
from the region of highest intensity, resulting in a repulsive potential.
Although vortices were not directly
imaged, significant heating of the cloud was observed only above a critical 
velocity, indicating a transition to a dissipative regime. This
heating was found to depend upon the existence of a condensate, indicating
that it must be due to the production of elementary 
excitations, that subsequently populate the 
non-condensed fraction.

Critical velocities for vortex formation in superflow past an obstacle 
have been studied numerically by solution of 
the GP equation in a homogeneous condensate \cite{frisch92,heupe97,winiecki99}.
Simulations have also confirmed that vortices are nucleated when a laser beam 
is translated inside a trapped Bose condensed gas \cite{jackson98,caradoc99}. 
In this paper, we attempt to clarify the role of vortices 
in the MIT experiment \cite{raman99} by presenting 2D and 3D simulations of
an oscillating repulsive potential in a
condensate.
The motion transfers energy to the condensate, and we observe that the
transfer rate increases significantly above the critical velocity for vortex
formation.

Our simulations employ the GP equation for the
condensate wavefunction  $\Psi (\mbox{\boldmath $r$,t})$ in a harmonic trap
$V_{\rm trap} (\mbox{\boldmath $r$}) = \frac{m}{2} \sum_{j} \omega_j^2 j^2$;
$j = x,y,z$. For convenience, we scale in harmonic oscillator units
(h.o.u.), where the units of length, time, and energy are 
$(\hbar/2m\omega_x)^{1/2}$, $\omega_x^{-1}$, and $\hbar \omega_x$, 
respectively. The scaled GP equation is then
\begin{equation}
 i \partial_t \Psi = (-\nabla^2+ V + C|\Psi|^2) \Psi,
\label{eq:GP-scaled}
\end{equation}
where $V$ represents a time-dependent `object' potential 
superimposed upon a stationary trap: $V=\frac{1}{4}(x^2+\epsilon y^2 +
\eta z^2)+V_{\rm ob} (\mbox{\boldmath $r$,t})$. The atomic interactions are
parameterized by $C=(NU_0 /\hbar \omega_x)(2m\omega_x/\hbar)^{\gamma/2}$: $N$
atoms of mass $m$ interact with a $s$-wave scattering length $a$, such that
$U_0=4\pi \hbar^2 a/m$. The number of dimensions is $\gamma$.
For most of the simulations here, $\gamma=2$, corresponding to   
the limit $\eta \rightarrow 0$. In this case, $N$
represents the number of atoms per unit length along $z$.
For the general 3D situation, a moving laser beam focused to a waist 
$\tilde{w}_0$ (in h.o.u.) at ($0,y'(t),0$), is simulated using
\begin{equation}
 V_{\rm ob} (\mbox{\boldmath $r$,t}) = \frac{U_{\rm ob}}{\sigma} \exp \left[
 \frac{-2(x^2+(y-y'(t))^2)}{\sigma \tilde{w}_0^2} \right],
\label{eq:ob-potential}
\end{equation}
where $\sigma=1+(z/z_0)^2$. The Rayleigh range is 
$z_0=\pi \tilde{w}_0^2/\lambda$, where  $\lambda$ is the laser wavelength
\cite{adams97}. 

Our numerical methods are discussed elsewhere 
\cite{jackson98,jackson00}. 
Briefly, initial states
are found by propagating (\ref{eq:GP-scaled}) in imaginary time with 
$V_{\rm ob} (\mbox{\boldmath $r$,0})$, using a spectral method. Then, 
real-time simulations are 
performed subject to motion of the object potential.
To recover the essential physics behind the MIT experiment \cite{raman99}, we
describe the oscillatory motion by $y'(T)=\alpha-vT$ ($T<1/2f$) and
$y'(T)=vT-3\alpha$ ($1/2f<T<1/f$), where $T=t-s/f$ and $s$ is the number of 
completed oscillations. 
The velocity between the motion extrema is constant,  
$\mbox{\boldmath $v$}=\pm 4\alpha f \mbox{\boldmath $\hat {y}$}$, where 
$\alpha$ is the amplitude and $f$ is the frequency. 
The condensate is anisotropic, with its long axis
along $y$ ($\epsilon < 1$). As a consequence, for small $\alpha$, the beam 
moves through regions of near-constant density.
Initially, the object 
creates a density minimum at $y=\alpha$, which follows closely behind the
moving object. For $v>v_c$, where $v_c \propto c_s$ and 
$c_s=\sqrt{2C|\Psi|^2}$ is the sound velocity, the density inside the beam 
evolves to zero. This is accompanied by a $\pi$ 
phase slip \cite{jackson98}, at which point the density minimum splits into a 
pair of vortex lines of equal but opposite circulation \cite{vortform}. The 
vortex pair separates, and the process begins again.

The creation of phonons or vortices increases the energy of the 
condensate, which was calculated numerically using the functional
$E = \int (|\nabla \Psi|^2 + V |\Psi|^2 + 
\frac{C}{2} |\Psi|^4 ) d^3 r$.
The time-independent ground state of the wavefunction represents the minimum 
of this functional. The energy is related
to the drag force on the object $\mbox{\boldmath $F$}_{\rm ob}$ by 
$dE/dt=\mbox{\boldmath $F$}_{\rm ob} . \mbox{\boldmath $v$}$. The drag
can be calculated independently over the whole condensate using 
$\mbox{\boldmath $F$}_{\rm ob} = -\int |\Psi|^2 \nabla{V}_{\rm ob} d^3 r$, 
allowing a numerical check.
Superfluidity corresponds to the situation where $E$
remains constant when $V_{\rm ob}$ is time-dependent; i.e.\ 
when there is no drag on the object.

\begin{figure}
\centering\epsfig{file=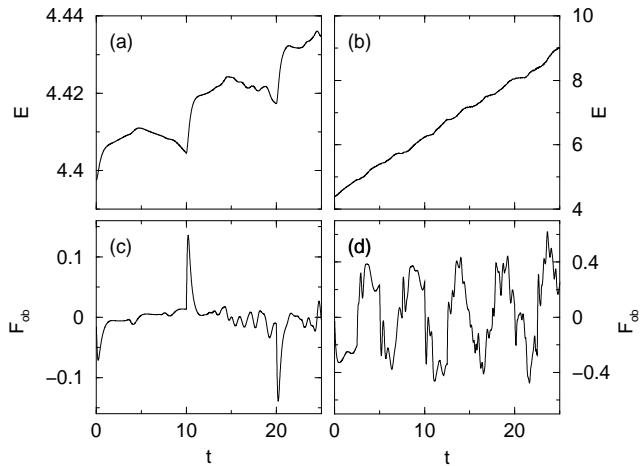, clip=, width=6.5cm, height=8.5cm,
 angle=270}
\caption{Time-dependent 2D simulations of laser beam oscillation, with 
 grid spacing of 0.156 ($512 \times 128$ points) and parameters $C=1000$, 
 $\epsilon=0.0625$, $\alpha=4$, $U_{\rm ob}=20$ and 
 $\tilde{w}_0=1.0$. Condensate energy as a function of time are
 plotted for (a) $f=0.05$ and (b) $f=0.2$. The drag $F_{\rm ob}$ is also 
 plotted for both frequencies in (c) and (d) respectively.}
\label{fig:Etransfer}
\end{figure}

\begin{figure}
\centering\epsfig{file=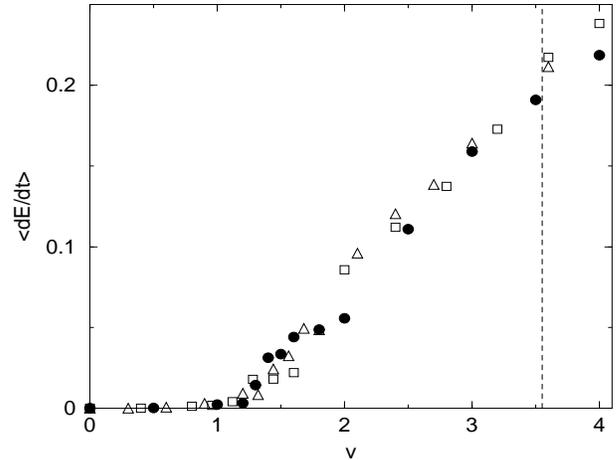, clip=, width=6.25cm, height=8.0cm,
 angle=270}
\caption{Mean rate of energy change as a function of velocity, for $\alpha=3$ 
 (triangles), $\alpha=4$ (squares) and $\alpha=5$ (bullets). Otherwise, 
 parameters are the same as Fig.\ \ref{fig:Etransfer}. The dashed line shows
 the speed of sound in the condensate center, $c_s=\sqrt{2\mu} \simeq 3.55$.
 The plot shows a sharp transition between phonon heating (low $v$)
 and vortex heating at $v_c \simeq 0.4 c_s$.}
\label{fig:transcurve}
\end{figure}

Fig.\ \ref{fig:Etransfer} shows the energy and drag as a function of time, as 
calculated for two different frequencies in 2D simulations. At low frequency, 
the energy transfer is relatively small and characterized by `jumps' at
the motion extrema, whereas at higher frequency the energy transfer is two 
orders of magnitude larger and more continuous. Further insight can be gained
by considering the drag. At low $f$, there is little drag
except at the motion extrema (Fig.\ \ref{fig:Etransfer}(c)), while at high 
$f$ appreciable drag is observed at all times (Fig.\ \ref{fig:Etransfer}(d)).
To measure the average rate of energy transfer, a
linear regression analysis is performed on the energy-time data. The gradients
are plotted against $v$ in Fig.\ \ref{fig:transcurve}. It can be
seen that the curves are characterized by two different regimes. Small energy
transfer at low $v$, gives way to enhanced heating above the
critical velocity, $v_c$. At high $v$, the three plots follow a single linear
curve.
Energy transfer below $v_c$ arises due to emission of sound waves at
the motion extrema. This process (henceforth referred to
as phonon heating)
is found to approximately scale with $v^3$, indicating that at each extremum 
(which are reached at a rate $\propto v$) a sound wave with energy 
$\sim v^2$ is emitted. Note phonon emission
by the object is not inconsistent with Landau's criterion. In particular, the
Landau argument relies on use of Galilean invariance, which breaks
down when the condensate density varies, or when the velocity changes abruptly.

For the parameters we have explored, phonon heating is found to be relatively
small compared to the energy transfer from vortex formation above $v_c$. The
heating rate in the latter regime is found to scale approximately linearly
with $v$. This implies that the drag force is constant. Indeed, we observe
that the drag saturates as $v$ increases. This behavior contrasts with
that of steady flow, where the drag
$\propto v^k$ (where $k \sim 1$ at $v$ close to $v_c$, and $k \rightarrow 2$
for $v>c_s$) \cite{frisch92,winiecki99}. The difference arises from the 
oscillatory motion: as the
object travels back through its own wake, a large pressure imbalance
across the object does not develop.

\begin{figure}
\centering\epsfig{file=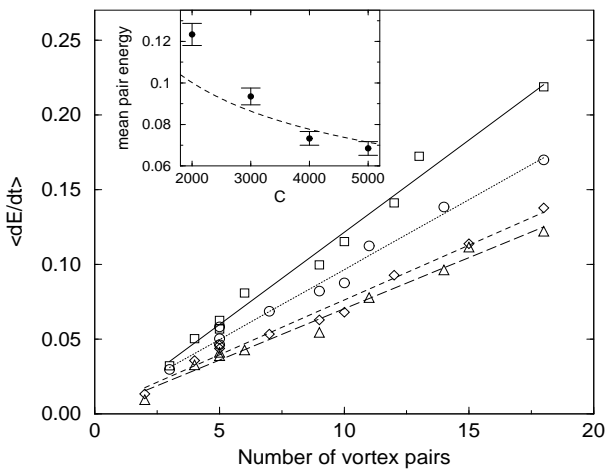, clip=, width=6.25cm, height=8.0cm,
 angle=270}
\caption{Number of vortex pairs created up to $t=10$ against rate of energy
 transfer. Simulation parameters are the same as Fig.\ \ref{fig:Etransfer}, 
 with $\alpha=4$ and $C=2000$ (plotted with squares, fit with a  solid
 linear regression line); $C=3000$ (circles, dotted line); $C=4000$ (diamonds,
 dashed); and $C=5000$ (triangles, long-dashed). The data points closely
 follow the regression lines, suggesting a constant energy for each vortex 
 pair. Inset: the average pair energy against $C$, where the dashed line 
 shows the pair energy predicted by (\ref{eq:pairE}).}
\label{fig:pairE}
\end{figure}

Fig.\ \ref{fig:pairE} plots the mean energy transferred against the number of
vortex pairs (counted in the simulated wavefunction). The energy transfer
per vortex pair is
approximately constant, leading to an estimate of the pair energy, which is
plotted as a function of nonlinearity
in Fig. \ref{fig:pairE} (inset). The energy of a vortex pair in an
homogeneous condensate with number density $n$, is given by
\begin{equation}
 E_{\rm pair} = \frac{2\pi n \hbar^2}{m} \ln \left(\frac{d}{\xi} 
 \right),
\label{eq:pairE}
\end{equation}
where $\xi$ is the healing length and $d$ is the distance between the vortices.
Equation (\ref{eq:pairE}) is valid for the inhomogeneous condensate when 
$\xi \ll d \ll R$, where $R$ is the radial extent of the condensate. Eq.\  
(\ref{eq:pairE}) with $d=2w_0$ is plotted in Fig.\ \ref{fig:pairE} (inset), 
and is found to agree with the numerical data.
Recall that the vortex pair separates
immediately after formation, when the pair still resides within the density
minimum created by the object. The pair also moves in the direction of the
object motion: however, it is slower, and is eventually left behind.
At this point, it has an energy approximately equal to $E_{\rm pair}$,
and the formation process is complete. The heating rate can be expressed
as $ dE/dt = E_{\rm pair} f_s$ \cite{raman99}, where $f_s$ is
the shedding frequency, which is found to be proportional to $v$.
This accounts for the linear dependence of the energy transfer rate.
The subsequent vortex dynamics involve an interplay between velocity fields
induced by other vortices, and effects arising from the condensate
inhomogeneity.
In the absence of the object, an isolated pair follows a trajectory
similar in character to that of a vortex ring \cite{jackson00}, culminating 
in self-annihilation. However, the
object moves back through its wake, interacting with the original pairs and
creating more vortices. The circulation of
a pair depends upon the direction of the object motion when it is created. 
So, vortex pairs of opposite circulation are formed and
interact when
sufficiently close. This leads to situations where vortices annihilate
or move towards the edge. The number of vortices
remaining within the condensate bulk is found to reach an equilibrium value. 
\begin{figure}
\centering\epsfig{file=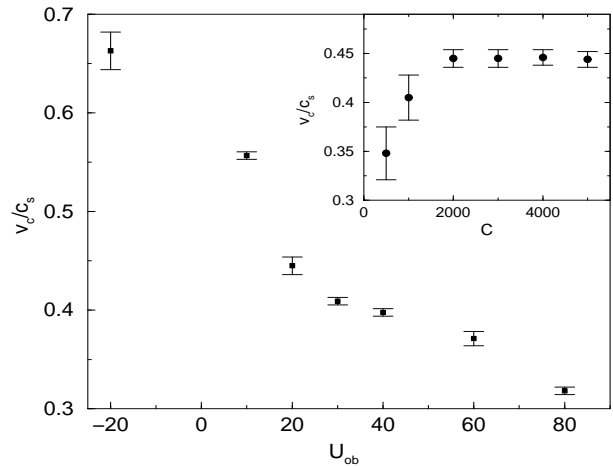, clip=, width=6.25cm, height=8.0cm,
 angle=270}
\caption{Critical velocity for vortex formation $v_c$ at $C=2000$ as a 
 function of 
 potential height $U_{\rm ob}$, expressed as a fraction of the speed of sound 
 at the
 condensate center, $c_s$. Inset: critical velocity plotted against $C$, with 
 $U_{\rm ob}=20$. The other parameters in both plots are $\alpha=4$ and
 $w_0=1$.}
\label{fig:crit-vel}
\end{figure}
The critical velocity for vortex formation, $v_c$, as a function of potential 
height
and nonlinear coefficient is shown in Fig.\ \ref{fig:crit-vel}. The critical 
velocity is not as 
well defined as in the
homogeneous case \cite{frisch92,heupe97,winiecki99} for a number
of reasons. First, a density inhomogeneity along the direction of motion
leads to a variation in $c_s$, and therefore $v_c$. However, this is less
than $\sim 3\%$ in the simulations considered here. The oscillatory nature
of the object motion is important. The time taken for a vortex pair to
form diverges to infinity as $v$ approaches $v_c$ from above. So, the measured
value of $v_c$ increases from its true value as $\alpha$ decreases. In 
addition, the object travels through its own low-density wake, where $c_s$
is lower. Vortices can therefore be formed after the first half-oscillation, 
when $v$ is slightly below $v_c$. Nevertheless, we can obtain a good estimate 
for $v_c$ by choosing
intermediate values of the amplitude (e.g.\ $\alpha=4$) and considering only
vortex formation during the first half-cycle.

Fig.\ \ref{fig:crit-vel} demonstrates that $v_c$ decreases as a function of 
increasing object potential height, $U_{\rm ob}$, allowing an experimental 
diagnostic for vortex formation at varying
beam intensities. This behavior agrees with simulations of 1D soliton
creation \cite{hakim97} and vortex ring formation in 3D \cite{winiecki99b}.
We have also studied the case of $U_{\rm ob}<0$, which 
corresponds 
to a red-detuned laser. Atoms are attracted to the potential minimum,
creating a density peak which moves with the beam. Vortex pairs are created
from a density minimum which develops ahead of the beam.
Fig.\ \ref{fig:crit-vel} (inset) shows $v_c$ as a function of
$C$. The critical velocity tends to a constant value as $C$ increases.
\begin{figure}
\centering\epsfig{file=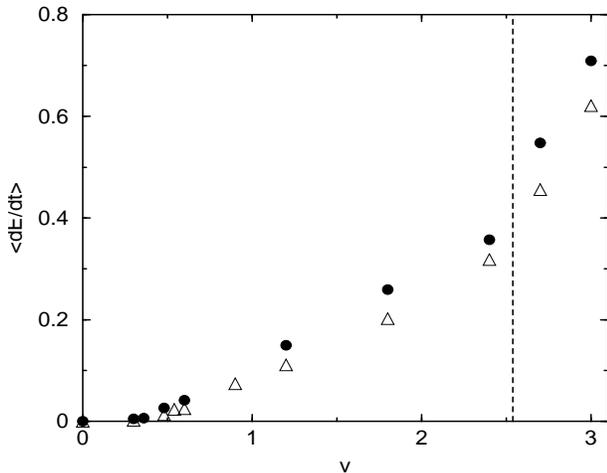, clip=, width=6.25cm, height=8.0cm,
 bbllx=91, bblly=49, bburx=576, bbury=626, angle=270}
\caption{Mean rate of energy change versus velocity, for 3D simulations with
 grid spacing 0.234 ($256\times 64 \times 64$ points). 
 Parameters are $C=1000$, $\epsilon=0.0625$, $\eta=1$, $\alpha=3.0$, 
 $\tilde{w}_0=1$, and 
 $\lambda \simeq 0.281$. The speed of sound at the condensate 
 center $c_s \simeq 2.54$ is represented 
 by the dashed line. For $U_{\rm ob}=40$ (bullets) the critical velocity is 
 $v_c \simeq 0.13 c_s$, while for $U_{\rm ob}=20$ (triangles) it is 
 $v_c \simeq 0.20 c_s$.}
\label{fig:threed}
\end{figure}
Simulations in 3D were performed, and the mean energy transfer rate as a 
function of velocity is presented in Fig.\ \ref{fig:threed}. Similar 
behaviour to 2D is observed, with smaller critical velocities: a result of the 
beam intersecting the condensate edge where 
the speed of sound is lower. Accordingly, vortex lines first appear in these
regions and penetrate into the center.
This conclusion agrees with the 
experiment \cite{raman99,numer}, where a relatively low critical velocity 
($v_c \simeq 0.26 c_s$) was measured. The dependence of $v_c$ on $U_{\rm ob}$
and $C$ was found to be similar to 2D, where e.g.\ $v_c \sim 0.29 c_s$ for
$C=4000$ and $U_{\rm ob}=35$. Enhanced heating is also observed for 
$v_c>c_s$, due to phonon emission between the extrema.

In this paper, we have studied the role of vortex formation in the breakdown
of superfluidity, by an oscillating object
in a trapped Bose-Einstein condensate. We find that at low object 
velocities, energy is transferred by phonon 
emission at the motion extrema, while a much larger energy is transferred 
above the critical 
velocity for vortex formation. To generalize these conclusions to realistic
experimental situations, the model should include the non-condensed thermal
cloud. Energy would then be transferred from the condensed to the thermal
cloud by phonon damping \cite{dalfovo99}, or vortex decay 
\cite{fedichev99}.

We acknowledge financial support from the EPSRC.


\begin{references}
\bibitem{vortform}
 Note that in general, the requirement that vortex pairs are formed `at a 
 point' follows from Kelvin's theorem, in order to conserve circulation
 within an arbitrary closed loop. For infinite potentials 
 \cite{frisch92,winiecki99}, vortex pairs are
 nucleated at the surface on opposite sides of the object, so that the 
 circulation within a loop intersecting the surface is not conserved. However,
 this loop is not closed, as fluid particles cannot enter into the object. 
 Hence, Kelvin's theorem is not violated in this case.
\bibitem{numer}
 Note that to simulate the experiment \cite{raman99} directly, the appropriate 
 parameters would be: $\epsilon \simeq 0.077$, $\tilde{w}_0
 \simeq 3.5$, $U_{\rm ob} \simeq 220$ and $C \simeq 3.6 \times 10^5$. 
 Such simulations require large grid sizes, making them computationally 
 intensive.
\end{references}
\end{document}